\baselineskip=12pt
\magnification 1200
\vsize=8.5 truein
\hsize=5.8 truein
\hoffset=1.0 cm
\overfullrule = 0pt

\font\ftit=cmbx10
\parskip=6pt
\parindent=2pc
\font\titulo=cmbx10 scaled\magstep1


\def\section#1{\vskip 1.5truepc plus 0.1truepc minus 0.1truepc
	\goodbreak \leftline{\titulo#1} \nobreak \vskip 0.1truepc
	\indent}
\def\frc#1#2{\leavevmode\kern.1em
	\raise.5ex\hbox{\the\scriptfont0 $ #1 $}\kern-.1em
	/\kern-.15em\lower.25ex\hbox{\the\scriptfont0 $ #2 $}}





\newbox\pmbbox
 \def\pmb#1{{\setbox\pmbbox=\hbox{$#1$}%
\copy\pmbbox\kern-\wd\pmbbox\kern.3pt\raise.3pt\copy\pmbbox\kern-\wd\pmbbox
\kern.3pt\box\pmbbox}}


\font\cmss=cmss10
\font\cmsss=cmss10 at 7pt

\def\IZ{\relax\ifmmode\mathchoice
{\hbox{\cmss Z\kern-.4em Z}}{\hbox{\cmss Z\kern-.4em Z}}
{\lower.9pt\hbox{\cmsss Z\kern-.4em Z}}
{\lower1.2pt\hbox{\cmsss Z\kern-.4em Z}}\else{\cmss Z\kern-.4em Z}\fi}

\font\cmss=cmss10
\font\cmsss=cmss10 at 7pt
\def\IS{\relax\ifmmode\mathchoice
{\hbox{\cmss S\kern-.4em S}}{\hbox{\cmss S\kern-.4em S}}
{\lower.9pt\hbox{\cmsss S\kern-.4em S}}
{\lower1.2pt\hbox{\cmsss S\kern-.4em S}}\else{\cmss S\kern-.4em S}\fi}


\parindent 0pt

\centerline{\titulo THE ASYMPTOTIC NUMBER OF ATTRACTORS}
\centerline{\titulo IN THE RANDOM MAP MODEL}

\vskip 1.2pc
\centerline{David Romero \ and \ Federico Zertuche}
\vskip 1.2pc
\centerline{Instituto de Matem\'aticas, UNAM}
\centerline{Unidad Cuernavaca, A.P. 273-3}
\centerline{62251 Cuernavaca, Morelos, Mexico.}
\centerline{\tt david@matcuer.unam.mx \ \ zertuche@matcuer.unam.mx}

\vskip 2.2pc

\centerline {\ftit Abstract}  
{\leftskip=1.5pc\rightskip=1.5pc The random map model is a
deterministic dynamical system in a finite phase space with $n$
points.  The map that establishes the dynamics of the system is
constructed by randomly choosing, for every point, another one as
being its image. We derive here explicit formulas for the statistical
distribution of the number of attractors in the system. As in related
results, the number of operations involved by our formulas
increases exponentially with $n$; therefore, they are not directly
applicable to study the behavior of systems where $n$ is large.
However, our formulas lend themselves to derive useful asymptotic
expressions, as we show.}
\vskip 1.2pc

\noindent Short Title: {\it The Asymptotic Number of Attractors}

\vskip 1.2pc


\


\vfill\eject

\section{1. Introduction}    

Since the 70's the random map model has attracted the attention of
physicists~$^{[1,2]}$. Indeed, in 1987 Derrida and Flyvbjerg have
shown that the random map model is equivalent to the Kauffman model of
cellular automata when the number of connections among the automata
goes to infinity~$^{[3,4]}$. This enlarged its application in the
realm of theoretical biology, disordered systems, and cellular
automata, for possible approaches of DNA replication, cell
differentiation, and evolution theory~$^{[1]}$.

On the other hand, one half of a century ago, some mathematicians had
approached the random map model in the context of random graphs.
First, in 1953, Metropolis \& Ulam posed the problem of determining the
number $\Theta(n)$ of expected connected components (i.e., attractors)
in random graphs with $n$ nodes~$^{[5]}$; at the time, $\Theta(n)$ was
estimated to be of order $\log n$.  Only one year later, Kruskal
elegantly solved the problem obtaining an exact formula together with
its corresponding asymptotic behavior~$^{[6]}$:
$$\Theta(n) = \sum_{k=1}^n {n! \over \left( n - k \right)! k n^k},
\eqno(1.1)$$ 
$$\Theta(n)\approx {1 \over2} \left( \ln 2 n + \gamma \right) +
\epsilon, \ {\rm for}\ n>>1, \eqno(1.2)$$ 
where $\epsilon$ vanishes for $n \to \infty$, and $\gamma$ is
the Euler-Mascheroni constant.

The statistical distribution of the number of connected components was
addressed both by Rubin \& Sitgreaves~$^{[7]}$, and Folkert in his
Ph.D.~thesis under Leo Katz supervision~$^{[8]}$.  Unfortunately,
these two practically out of reach works were never published even
though their relevance and an offer by Katz to do so~$^{[9]}$. 
Later on, Harris partially reviewed and enlarged these results,
proposing a new combinatorial expression for the distribution of
connected components, as well as the complete solution of the simpler
case in which the random map is one-to-one~$^{[10]}$.  The
mathematical expressions of a general random map found by Folkert and
Harris, however correct, are based on constrained sums with a number
of terms of order e$^n$, and involve Stirling numbers of the first
kind~$^{[11-13]}$. Therefore, unfortunately, their straightforward use
in physics or biology appears quite limited (for example, it is
typical to deal with $n\sim 2^{100}$ in models such as those of
cellular automata).

In spite of its importance, and as far as we know, 
a study of the variance for the distribution of connected components 
has not yet been undertaken. 

In this paper we propose a still new combinatorial formula
---equivalent to the previous ones, of course--- for the statistical
distribution of the number of connected components.  As in earlier
results, it also relies on a constrained sum, and the involved
computational effort increases exponentially with $n$.  Nevertheless,
it has the advantage of allowing us the derivation of the long-needed
asymptotic formula for the statistical distribution.  Furthermore, we
easily deduce from it asymptotic formulas for the corresponding
average and variance.

The paper is organized as follows.  Section~2 is devoted to the main
definitions of the random map model, and to settle our conventions.
Then, in Section~3, we determine an exact combinatorial expression for
the statistical distribution of the number of connected components in
the model.  The corresponding asymptotic formula for this distribution
is derived in Section~4, as well as asymptotic formulas for the
average and the variance for the number of connected components.  To
end, we present our conclusions in Section~5.

\section{2. The Random Map Model and Functional Graphs}

Let $\Omega = \{1,2,\dots,n\}$ be a set of $n$ points.  To each point
in $\Omega$ assign at random one point in $\Omega$ with uniform
probability distribution, thus defining a function
$f:\Omega\rightarrow\Omega$.  
In this way a dynamical system has been
established on the so-called phase space $\Omega$ through the
iterations of $f$; this is the random map model~$^{[3,4]}$.

Since $\Omega$ is finite every orbit of $f$ will eventually end in a
periodic attractor, and several questions are in order, like:
what is the expected number of attractors in the system? which is the
statistical distribution of the number of attractors? how large is
the dispersion of the distribution?  Here, we answer these questions
starting from combinatorial arguments.

For each function $f$ on $\Omega$ define a {\it functional graph}
whose nodes are precisely the elements of $\Omega$; moreover, if
$f(i)=j$ then draw a directed link from node $i$ to node $j$, and
whenever $f(i)=i$ a loop on node $i$ is drawn.  As an example,
Figure~1 shows a functional graph with three connected components
(i.e., three attractors) in a set with $n=11$.

Note that each function $f$ on $\Omega$ (or functional graph) can be
represented by an $n\times n$ binary matrix $M=\{M_{ij}\}$, where
$M_{ij}=1$ whenever $f(i)=j$ and $M_{ij}=0$ otherwise.  Every row of
matrix $M$ has $n-1$ zeros and one `1'.  Clearly, there exist $n^n$
such matrices, and thus $n^n$ is the number of functions $f$ that can
be defined on $\Omega$, as well as the number of distinct functional
graphs on $n$ nodes.

\midinsert
\vbox to 3truein{\includegraphics{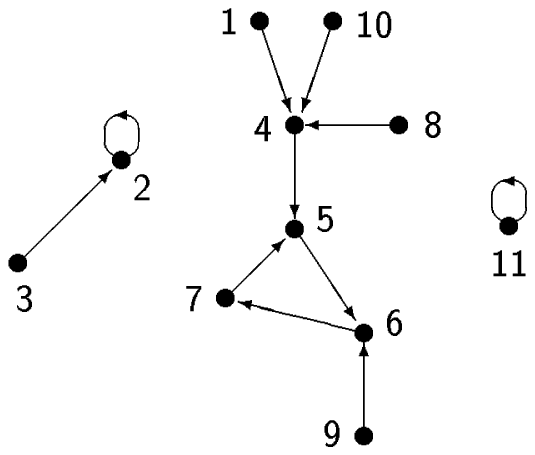}}
\noindent \item{[1]} A functional Graph with three connected
components for $ n = 11 $.
\endinsert

\vfill \eject

\

\

\section{3. The Distribution of Connected Components} 

What is the number $a_n$ of connected functional graphs (i.e., having
precisely one connected component) that can be found among the $n^n$
functional graphs on $n$ nodes?  Through clever combinatorial
arguments Katz obtained~$^{[9]}$
$$a_n=(n-1)!\sum_{k=0}^{n-1}{n^k\over k!}, \eqno(3.1)$$
and showed that, for $n>>1$, this figure grows asymptotically like
$$a_n \approx n^n \sqrt{\pi \over 2n} \left(1 + {\cal O}
\left( {1 \over n} \right) \right). \eqno(3.2) $$

Using (3.1) we derive now a new expression for the statistical
distribution of the number of connected components in functional
graphs.  Consider first a partition ${\cal P}$ of $\Omega$, in $k$
disjoint subsets $\Omega_1,\dots,\Omega_k$, with $n_1,\dots,n_k$
points, respectively; then
$$\sum\limits_{j=1}^k n_j=n,\hskip14pt\hbox{with}\hskip4pt 1\leq n_j\leq n, 
\hskip3pt\hbox{for}\hskip4pt j=1,\dots,k. \eqno(3.3a)$$

Obviously, $\prod_{j=1}^k a_{n_j}$ is the number of functional graphs
$G$ with $k$ connected components, such that the subset $\Omega_j$
defines a connected component of $G$, for $j=1,\dots,k$.  Moreover,
for given integers $n_1,\dots,n_k$, satisfying (3.3a), the multinomial
coefficient ${n\choose n_1,\dots,n_k} = {n!\over n_1!\cdots n_k!}$
yields the number of distinct ways we can distribute $n$ objects in
distinguishable boxes $B_1,\dots,B_k$, of sizes $n_1,\dots,n_k$,
respectively~${}^{[11,13]}$. This leads to
	$${1\over k!}\sum\limits_{\{n_1,\dots,n_k\}'} 
	{n\choose n_1,\dots,n_k}\prod_{j=1}^k a_{n_j},$$
as the number of functional graphs with $k$ connected components,
where $\{n_1,\dots,n_k\}'$ means that the sum is over all vectors 
$(n_1,\dots,n_k)$ satisfying (3.3a), and the factor ${1\over k!}$ 
stands to account for the unavoidable repetitions incurred by our above 
assumption of `distinguishable boxes'.

Therefore, for $k=1,\dots,n$, the distribution for the number of 
connected components can be expressed as
$$\rho_n(k)={1\over n^n k!}\sum\limits_{\{n_1,\dots,n_k\}'} 
{n\choose n_1,\dots,n_k}\prod_{j=1}^k a_{n_j}. \eqno(3.3b) $$

It is not difficult to see that the sum in (3.3b) has as much as
${n-1\choose k-1}$ terms~${}^{[13]}$, 
a figure growing exponentially with $n$ when 
$k\sim{n\over2}$.
This fact dooms to failure any numerical application of (3.3b) as it 
stands, and hinders further analytical work. 

The approaches of Folkert and Harris lead to increasing complications.
The former yields~$^{[8,10]}$
$$\rho_n(k) = {1\over n^n} \sum\limits_{\mu=k}^n
{{\cal S}_\mu^{(k)} \over \mu!} \sum\limits_{\{n_1,\dots,n_\mu\}''} 
{n \choose n_1,\dots,n_\mu} n_1^{n_1}\dots
n_\mu^{n_\mu}, \eqno(3.4a) $$
where ${\cal S}_\mu^{(k)}$ are the Stirling numbers of
the first kind~$^{[11-13]}$ (see Appendix), and $\{n_1,\dots,n_\mu\}''$ means that
the sum is over all vectors $n_1,\dots,n_\mu$, constrained by
$$ \sum\limits_{j=1}^\mu n_j = n \hskip 11pt \hbox{with}
\hskip 9pt 1 \leq n_j \leq n. \eqno(3.4b) $$
 
Harris managed to propose~$^{[10]}$ 
$$ \rho_n \left( k \right) = {n!\over n^n k!} \sum\limits_{\mu=k}^n
{\cal S}_\mu^{\left( k \right)} \sum\limits_{\{n_1,\dots,n_n\}'''} 
{1 \over n_1!,\dots,n_n!} \left( {1^1 \over 1!} \right)^{n_1} 
\left( {2^2 \over 2!} \right)^{n_2} \dots \left({n^n \over n!}\right)^{n_n}, 
\eqno(3.5a) $$
in which $\{n_1,\dots,n_n\}'''$ means that the sum is over all vectors 
$n_1,\dots,n_n$, constrained by
	$$ \sum\limits_{j = 1}^n n_j = \mu \hskip 9pt \hbox{and}
	\hskip 9pt \sum\limits_{j = 1}^n j n_j = n \hskip 11pt \hbox{with}
	\hskip 9pt 0 \leq n_j \leq n. \eqno(3.5b) $$

Expressions (3.4) and (3.5) appear more difficult to handle than (3.3)
because of the added extra terms via the summation involving 
Stirling numbers. 

To our knowledge, the statistical moments of the distribution for the
number of connected components have not yet been obtained in closed
form. Further, the exact calculation done by Kruskal for the expected
number of attractors (1.1) is not derived from a distribution (see
Ref.~[6]).  For application purposes ($n>>1$) it is then worthwhile to
derive manageable asymptotic formulas for $\rho_n(k)$; this is the
subject of the next section.

\section{4. Asymptotic Expressions}

Let us start by defining
$$\beta_m=2 e^{-m} {a_m\over\left(m-1 \right)!}\eqno(4.1)$$
and 
	$$\alpha_m = 1 - \beta_m. \eqno(4.2) $$ 

Due to the asymptotic relation (3.2) it happens that 
$$\beta_m = 1 + {\cal O} \left({1 \over m} \right),\hskip 14pt 
\hbox{for} \hskip 6pt m >> 1, \eqno(4.3a) $$
and
	$$ \alpha_m = {\cal O} \left({1 \over m} \right), \hskip 14pt
	\hbox{for} \hskip 6pt m >> 1. \eqno(4.3b) $$

Now, express (3.3b) in terms of $\beta_m$ and use (3.3a) to get  
$$\rho_n\left(k\right) = {n! e^n \over n^n k! 2^k}\sum\limits_{\{n_1,\dots,n_k\}'}
\prod_{j=1}^k {\beta_{n_j}\over n_j}.$$ 

Constraint (3.3a) can be broken by introducing a Kronecker delta
inside the summation. 
Using the integral representation
$$\delta_{n, m} = {1 \over 2 \pi i} \oint_\Gamma{z^m \over z^{n+1}}dz,$$
where $\Gamma$ is any closed contour in the complex plane of $z$
such that the origin is inside, we go back to the expression for 
$\rho_n(k)$ to obtain
$$\rho_n \left( k \right) = {n! e^n \over n^n k! 2^k 2 \pi i}
\oint_\Gamma  {dz \over z^{n+1}} g_k \left( z \right)$$ 
with a modicum of algebra, where 
$$g_k(z)= \left[ \varphi_p \left( z \right) \right]^k \hskip 11pt \hbox{with}
\hskip 9pt \varphi_p \left( z \right) = \sum\limits_{m=1}^p {\beta_m
\over m} z^m \hskip 11pt \hbox{and} \hskip 9ptp \geq n. \eqno(4.4a) $$

Since $g_k(z)$ is an analytic function we can apply the Cauchy
integral theorem to find
$$\rho_n(k) = {e^n\over n^n k! 2^k} g_k^{(n)}(0) \eqno(4.4b)$$
as an alternate way to compute $\rho_n(k)$. 

The computation of (4.4) may be as difficult as that of (3.3); however, the
former is manageable for $n>>1$, as we shall see. 
First note that (4.4b) does not depend on the particular value we give
to $p$ in (4.4a), as long as $p\geq n$, a feature that we shall use 
to our convenience. 
Taking $p=n$ in (4.4a), $g_k(z)$ becomes a polynomial of degree $nk$, 
and its $n$-th derivative may be calculated in terms of finite differences 
with the help of Stirling numbers (see formulas ($A$.6) in Appendix), 
obtaining
	$$ g_k^{\left(n\right)} \left(0\right)=n!
\sum\limits_{r=n}^{nk} {{\cal S}_r^{\left( n \right)} \over r!}
\Delta^r g_k \left(0\right), \eqno(4.5a) $$ 
where ${\cal S}_r^{(n)}$ are the Stirling numbers of the first kind,
and
$$\Delta^r g_k(0)=\sum\limits_{m=1}^r(-1)^{r-m}{r\choose
m}g_k(m). \eqno(4.5b) $$

Now we provide asymptotic approximations for $ \varphi_p \left( z
\right) $ in (4.4a) with $ p = n $ and $ n >> 1 $. Consider first $
\mid z
\mid > 1 $, and note that, due to (4.3),
	$$ \varphi_n \left( z \right) = \sum\limits_{k=1}^n {\beta_k
	\over k} z^k \approx \sum\limits_{k=1}^n {1 \over k} z^k \left( 1 +
	{\cal O} \left( {1 \over n} \right) \right). $$ 

Now, the Euler-Maclaurin formula states~$^{[11,14]}$
	$$ \eqalignno{\sum_{k=1}^n f \left( k \right) \approx &\int_1^n f \left( x
	\right) dx + {1 \over 2} \left( f \left( n \right) + f \left( 1
	\right) \right)  \cr &+\sum_{k=1}^\infty {B_{2k} \over \left( 2k
	\right) !} \left( f^{\left( 2k-1 \right) } \left( n \right) -
	f^{\left( 2k-1 \right) } \left( 1 \right) \right), &(4.6a)} $$
where $ f \left( x \right) $ is a $ C^\infty $ function over the
interval $ \left[ 1, n \right] $, and $ B_{k} $ are the Bernoulli
numbers defined by the generating function 
$${t \over e^t - 1} = \sum_{k=0}^\infty {B_k \over k!} t^k
\hskip 11pt \hbox{for} \hskip 9pt \mid t \mid < 2 \pi, \eqno(4.6b) $$
with $ B_{2k+1} = 0 $, for $k=1,2,\dots$ 

Let us take $f(x)={z^x\over x}$ with $\mid z\mid>1$, and substitute in (4.6a)
obtaining 
	$$ \sum\limits_{k=1}^n {\beta_k \over k} z^k \approx {z^n
	\over n} \left[ {1 \over \ln z} + {1 \over 2} +
	\sum_{k=1}^\infty {B_{2k} \over \left( 2k \right) !} \left(
	\ln z \right)^{2k-1} \right] \left( 1 + {\cal O} \left( {1
	\over n} \right) \right). $$
Using (4.6b) yields
$$ \sum\limits_{k=1}^n {\beta_k \over k} z^k \approx {z \over
z - 1} {z^n \over n} \left( 1 + {\cal O} \left( {1 \over n}\right) \right) 
\hskip 11pt \hbox{for} \hskip 9pt 1 < \mid z \mid < e^{2 \pi},$$
which can be continued analytically for any $\mid z\mid>1$,
thus obtaining
	$$ \varphi_n \left( z \right) = \sum\limits_{m=1}^n {\beta_m
	\over m} z^m \approx {z \over z - 1} {z^n \over n} \zeta_n \hskip 11pt
	\hbox{for} \hskip 9pt \mid z \mid > 1,\eqno(4.7a)
	$$ where
	$$ \zeta_n = 1 + {\cal O}\left( {1 \over n} \right). \eqno(4.7b)
	$$
Instead, for $ z = 1 $, clearly
	$$\varphi_n \left( 1 \right) = \sum\limits_{m=1}^n {\beta_m
	\over m} \approx \left( \ln n + \gamma - \tau_n \right) \eta_n,
	\eqno(4.7c) $$ where
	$$\eta_n = 1 + {\cal O} \left( {1 \over n \ln n }
\right), \eqno(4.7d) $$  
	$$\tau_n = \sum_{m=1}^n {\alpha_m \over m}, \eqno(4.7e) $$ 
and $\gamma = 0.577\ 215\ 66 \dots$ is the Euler-Mascheroni constant. 
Note that, due to (4.3b), when $n\rightarrow\infty$ in (4.7e), 
the limit $\tau$ exists. 
Substituting back (4.7a,b) into (4.4a), expanding then in Taylor series 
the term ${z \over z-1}$ for $\mid z\mid > 1$, and cutting the series 
beyond order $nk$ we come to
	$$ g_k \left( z \right) \approx {z^{nk} \over n^k} \zeta_n^k
	\sum_{m=0}^{nk} {\left( k \right)_m \over m!} {1 \over z^m}
	\left( 1 + {\cal O} \left( {1 \over z^{nk}} \right) \right)
	\hskip 11pt \hbox{for} \hskip 9pt \mid z \mid > 1, \eqno(4.8a) $$
with
$$(k)_m = k(k + 1)\dots(k+m-1) = {\Gamma(k+m) \over \Gamma(k)}. $$ 
While substituting (4.7c-e) into (4.4a) gives us
	$$ g_k \left( 1 \right) \approx \left( \ln n + \gamma - \tau_n
	\right)^k \eta_n^k. \eqno(4.8b) $$ 

Replace (4.8) into (4.5b) to get 
	$$ \Delta^r g_k \left( 0 \right) \approx \left( -1
	\right)^{r-1} r \left[ g_k \left( 1 \right) -
	\lambda_n^{\left( k \right)} \right] + {r! \over 
	n^k} \zeta_n^k \sum_{l=0}^{nk} {\left( k \right)_l \over l!}
	{\cal T}_{nk-l}^{\left( r \right)},  \eqno(4.9) $$
where we have used ($A$.4) to introduce the Stirling numbers of the
second kind ${\cal T}_p^{(l)}$, and 
$$\lambda_n^{(k)}\equiv{\zeta_n^k \over n^k}\sum_{l=0}^{nk}{(k)_l\over l!}.$$
 
The summation for $\lambda_n^{(k)}$ can be found in summation tables 
yielding~$^{[15]}$
$$\sum_{l=0}^{nk} {(k)_l \over l!} = 
{\left[k\left( n+1 \right) \right] ! \over k!(nk)!}.$$ 

Next, applying Stirling's approximation ($s! \approx \sqrt{2 \pi s} s^s
e^{-s} $ for $ s >> 1 $) to the factorials for $n >> 1$, one finds
$\lambda \sim e^k$, while $g_k(1)\approx e^{k\left\{\ln[\ln n + \gamma
- \tau_n]+\ln \eta_n \right\}}$; thus, $\lambda$ is exponentially
small as compared to $g_k\left(1\right)$, and can be neglected in
(4.9).  Replacing back (4.9) into (4.5a), using properties ($A$.5), and
substituting in (4.4b), we find for the asymptotic distribution of the
number of attractors
	$$ \rho_n \left( k \right) \approx {\sqrt{2\pi n} \over k!
	2^k} g_k \left( 1 \right) \sum_{l=n}^{nk} {\left( -1
	\right)^{l-1} \over \left( l - 1 \right) ! } {\cal
	S}_k^{\left( n \right) } + {\cal N}_n \left( k
\right), \eqno(4.10) $$
where 
	$$ {\cal N}_n \left( k \right) = {\sqrt{2\pi n} \over k! 2^k}
	{1 \over n^k } \zeta_n^k {\left( k \right)_{\left[n \left(k-1 
	\right)\right]} \over \left[ n \left( k - 1 \right)
	\right]!}. \eqno(4.11) $$
 
We show now that ${\cal N}_n(k)$ is a null term, that is to say 
$$ \lim\limits_{n \rightarrow \infty} \sum_{k=1}^n k^m {\cal
N}_n (k) = 0 \eqno(4.12)$$
for any $m\geq0$, and therefore it does not contribute to the averages taken
with $\rho_n(k)$, for $n>>1$. 
First note that ${\cal N}_n\left(k\right)\sim{1\over\sqrt{n}}$ for any $k$,  
so as in the summation (4.12) the terms with $k\sim{\cal O}(1)$ 
do not contribute. 
Hence, we can make the approximation for $n>>1$ and $k>>1$ in (4.11), 
and calculate (4.12) with it, obtaining 
	$$ {\cal N}_n \left( k \right) \approx {e^{-\left(1 + {1 \over
	2n} \right)} \over \sqrt{n}} {1 \over k!} \left[ {1 \over 2}
	\zeta_n e^{ 1 + {1 \over 2n}} \right]^k.$$

Since 
$$\sum_{k=1}^\infty k^m {x^k \over k!} = {\cal P}_m(x) e^x,$$
where ${\cal P}_m (x)$ is a polynomial in $x$ of degree $m$, we have
	$$ \sum_{k=1}^n k^m {\cal N}_n \left( k \right) \approx
	{e^{-\left(1 + {1 \over 2n}\right)} \over \sqrt{n}} {\cal P}_m
	\left( {1 \over 2} \zeta_n e^{1 + {1 \over 2n}} \right) \exp
	\left( {1 \over 2} \zeta_n e^{1 + {1 \over 2n}} \right),$$
and thus the expression goes to zero as ${1 \over \sqrt{n}}$, which
proves (4.12). 

Then, (4.10) can be expressed very simply in terms of Stirling
numbers of the first kind by noting the following: 
let us take for a moment $\alpha_m=0$ in (4.4a), and also set $p=\infty$ 
with $\mid z \mid < 1$ (so the infinite sum converges), obtaining
$$ \widetilde{g}_k \left( z \right) = \left( \sum_{m=1}^\infty
{1 \over m} z^m \right)^k = \left\{ - \ln \left( 1 - z \right)\right\}^k,$$
where the tilde indicates that we have set $\alpha_m = 0$. 
Using ($A$.2) and substituting in (4.4b) we arrive to 
	$$ \widetilde{\rho}_n \left( k \right) = {e^n \over n^n 2^k}
	\left( -1 \right)^{n-k} {\cal S}_n^{\left( k \right)}.$$

Comparing with (4.10), we obtain, up to the null term, that
	$$ \rho_n \left( k \right) \approx {e^n \over n^n} \left( -1
	\right)^{n-k} {\cal S}_n^{\left( k \right)} \left( {\mu_n
	\over 2 } \right)^k, \eqno(4.13a) $$ 
where 
	$$ \mu_n = 1 - {\tau_n \over \gamma + \ln n}. \eqno(4.13b) $$
 
Since $ \rho_n \left( k \right) $ is a statistical distribution, 
$$ \lim_{n \rightarrow \infty} \sum_{k=1}^n \rho_n(k) = 1.$$
Which allows us to compute $\tau=\lim_{n\rightarrow\infty}\tau_n$ 
in (4.13b), by using property ($A$.1) of Stirling numbers, giving 
$$ \tau = \ln 2. \eqno(4.13c) $$
 
We can replace back (4.13c) into (4.13a) and (4.13b), yielding and
error of order ${\cal O} \left( 1 / \ln n \right)$, for $ 1 \leq k
\leq n $.
So we have, for the asymptotic approximation of the distribution of the 
number of attractors, the following expression:
	$$ \rho_n \left( k \right) \approx {e^n \over n^n} \left( -1
	\right)^{n-k} {\cal S}_n^{\left( k \right)} \left( {\mu
	\over 2 } \right)^k \left( 1 + {\cal O} \left( {1 \over \ln n}
	\right) \right), \eqno(4.14a) $$ 
where
$$\mu = 1 - {\ln 2 \over \ln n}. \eqno(4.14b) $$ 

Now we can use (4.14) to calculate averages. 
Using expression ($A$.3), we have 
	$$ \Theta\left( n \right) \ \approx {1 \over2} \left( \ln 2 n
	+ \gamma + {\cal O} \left( {1 \over \ln n} \right) \right)
	\eqno(4.15) $$ 
for the average number of attractors $ \Theta\left( n \right) $, which
coincides, as it should be, with Kruskal asymptotic
approximation (1.2). The variance can also be computed by using
the derivative of ($A$.3):
	$$ \sigma^2\approx\ \Theta \left( n \right)  \left(1+{\cal
	O}\left({1\over\ln n}\right)\right), \eqno(4.16) $$
which is a new result.

\section{Conclusion}

We have proposed a new and simpler expression for the distribution
of the number of attractors in the random map model (Eq.~(3.3)). The
number of operations involved for the numerical evaluation of the
distribution grows exponentially with $n$, thus making expression
(3.3) useless for direct calculations for $ n>>1 $. To overcome this
difficulty, however, we have derived an asymptotic formula
(Eq.~(4.14)), from which we directly deduced asymptotic values both
for the average number of attractors and for its variance (Eqs.~(4.15)
and (4.16), resp.).

In the random map model, additional statistical figures are of
interest. Among them, the average attractor size, the average period 
(or average length) of the cycles; also, 
given a point $x$ in the phase space, the
expected length of its orbit, the expected length of the cycle in the
attractor containing $x$, the expected number of points from which $x$
can be attained, etc. 
Some of these figures have already been computed (for example, 
see Ref.~[2,3,10]), and others are subject of recent
studies~$^{[16]}$.

\section{Acknowledgments} 

This work is supported in part by {\bf CONACyT} project number
U40004-F.  The authors would like to thank V. Dom\'\i nguez,
E. Sacrist\'an, F. Toledo for computational advice. The second author
(FZ) thanks Alberto Verjovsky for his invaluable clarification of some
aspects of asymptotic series.

\vfill\eject

\section{Appendix: Properties of Stirling Numbers}

Here we list the equations and definitions on Stirling numbers
necessary to follow the calculations in this article. 
For a more extensive treatment see Refs.~[11-13].
Stirling numbers of the first kind are generated by the functions
$$z(z+1)\dots(z+n-1) = \sum\limits_{m=1}^n(-1)^{n-m} {\cal S}_n^{(m)} 
z^m \eqno(A.1)$$
and
$$\left\{ \ln \left( 1 + z \right) \right\}^k = k!
\sum\limits_{r=k}^\infty {{\cal S}_r^{(k)} \over r!} z^r 
\hskip 11pt \hbox{for} \hskip 9pt \mid z \mid < 1. \eqno(A.2) $$

Applying the operator $z{d\over dz}$ to equation ($A$.1) we get the
important relation
$$z \left( z + 1 \right) \dots \left( z + n - 1 \right)
\sum\limits_{m=1}^n {z \over z + m - 1} = \sum\limits_{m=1}^n
\left( -1 \right)^{n-m} {\cal S}_n^{\left( m \right)} m z^m. \eqno(A.3)
$$

Stirling numbers of the first kind may be expressed in closed form by 
$${\cal S}_k^{\left( n \right)} = \sum\limits_{l=0}^{k-n}
\left( -1 \right)^l {k-1+l \choose k-n+l} {2k-n \choose k-n-l}
{\cal T}_{k-n+l}^{\left( l \right)},$$

where
$${\cal T}_p^{\left( l \right)} = {1 \over l!}
\sum\limits_{k=0}^l \left( -1 \right)^{l-k} {l \choose k} k^p
\eqno(A.4)$$

are the Stirling numbers of the second kind.
It is known that Stirling numbers verify the relations
	$$ {\cal S}_k^{\left( n \right)} = 0 \hskip 11pt \hbox{if}
	\hskip 9pt k < n, \eqno(A.5a) $$
	$$ {\cal T}_p^{\left( l \right)} = 0 \hskip 11pt \hbox{if}
	\hskip 9pt p < l, \eqno(A.5b) $$
and
	$$ \sum\limits_{k=m}^n {\cal S}_k^{\left( m \right)} {\cal
	T}_n^{\left( k \right)} = \sum\limits_{k=m}^n {\cal
	S}_n^{\left( k \right)} {\cal T}_k^{\left( m \right)} =
	\delta_{m,n}. \eqno(A.5c) $$

By means of Stirling numbers of the first kind it is possible to
express derivatives of a function in terms of finite differences by
the formula
	$$ {d^m \over dz^m} f\left( z \right) = m!
	\sum\limits_{k=m}^\infty {{\cal S}_k^{\left( m \right)} \over
	k!} \Delta^k f \left( z \right), \eqno(A.6a) $$

if the summation is convergent, and where 
$\Delta f\left(x\right) = f\left(x+1\right) - f\left(x\right)$ and 
	$$ \Delta^k f \left( z \right) = \sum\limits_{l=0}^k \left( -1
	\right)^{k-l} {k \choose l} f \left( z + l
	\right). \eqno(A.6b)$$ 
	
\vfill\eject

\centerline{{\ftit References}}

\item{[1]} Kauffman, S.A., {\it Metabolic stability and
epigenesis in randomly connected nets}, J.~Theoret.~Biol. {\bf 22}
(1969) 437; {\it Cellular homeostasis, epigenesis and replication in
randomly aggregated macromolecular systems}, J.~Cybernetics {\bf 1}
(1971) 71; Wolfram, S., {\it Statistical mechanics of cellular
automata}, Rev.~Mod.~Phys. {\bf 55} (1983) 601; {\it Guanajuato
Lectures on Complex Systems and Binary Networks}. Springer Verlag
Lecture Notes series. Eds. R.  L\'opez Pe\~na, R. Capovilla,
R. Garc\'\i a-Pelayo, H. Waelbroeck and F. Zertuche. (1995); Aldana
M., Coppersmith S. and Kadanoff L. {\it Boolean Dynamics with Random
Couplings}. nlin.AO/020406 (2002).

\item{[2]} Coste, J.~and Henon, M., In {\it Disordered systems and 
biological organization}, Eds. Bienenstock, M.Y., Fogelman Souli\'e
F., and Weisbuch, G., p.~361, Springer Verlag, Heidelberg (1986).

\item{[3]} Derrida, B.~and Flyvbjerg, H., {\it The random map model: 
a disordered model with deterministic dynamics}, J.~Physique {\bf 48} 
(1987) 971.

\item{[4]} Kauffman, S.A., {\it The origins of order: Self-organization 
and selection in evolution}. Oxford University Press (1993).

\item{[5]} Metropolis, N.~and Ulam, S., {\it A property of
randomness of an arithmetical function}, Amer.~Math.~Monthly {\bf 60}
(1953) 252. 

\item{[6]} Kruskal, M.D., {\it The expected number of components under 
a random mapping function}, Am.~Math.~Monthly {\bf 61} (1954) 392. 

\item{[7]} Rubin, H.~and Sitgreaves, R., {\it Probability
distributions related to random transformations on a finite
set}, Tech.~Rept. No.~19A Applied Mathematics and Statistics
Laboratory, Stanford University (1954). Unpublished. 

\item{[8]} Folkert, J.E., {\it The distribution of the number
of components of a random mapping function}, (1955) Unpublished
Ph.~D. dissertation, Michigan State University. 

\item{[9]} Katz, L., {\it Probability of indecomposability of
a random mapping function}, Ann. Math.~Stat. {\bf 26} (1955) 512. 

\item{[10]} Harris, B., {\it Probability distributions related
to random mappings}, Ann.~Math. Stat. {\bf 31} (1960) 1045. 

\item{[11]} Abramowitz, M.~and Stegun, I.A., {\it Handbook of
mathematical functions}. Dover Publications, New York (1972). 

\item{[12]} Jordan, C., {\it Calculus of finite differences}.
Chelsea Publishing Company, New York (1947).

\item{[13]} Graham, R.L., Knuth, D.E. and Patashnik, O., {\it Concrete
Mathematics.} Addison-Wesley Publishing Company, New York (1994).

\item{[14]} Arfken, G., {\it Mathematical Methods for Physicists}.
Academic Press, New York (1970) Chap.~10.

\item{[15]} Mangulis, V., {\it Handbook of Series for Scientists and 
Engineers}. Academic Press. New York and London (1965). Page~60.

\item{[16]} Romero, D.~and Zertuche, F., {\it Grasping the
Connectivity of Functional Graphs}, submitted for publication (2002).

\end




\vskip 7cm

\

\

\midinsert
\vbox to 3truein{\includegraphics{figure1.ps}}
\vskip 0.5cm
\noindent \item{[1]} A functional Graph with three connected
components for $ n = 11 $.
\endinsert